\documentclass{elsart3}
\usepackage{graphicx}
\begin{document}

\begin{frontmatter}
\title{New limits on the ordered moments in $\alpha$-Pu and Ga-stabilized $\delta$-Pu}
\author[LANL,JAERI]{R.H.~Heffner\corauthref{cor}}
\corauth[cor]{Tel. +81-29-284-3524,
Fax: +81-29-282-5927,
email: heffner@popsvr.tokai.jaeri.go.jp} 
\author[LANL,TRI]{G.D.~Morris}
\author[LLNL]{M.J.~Fluss}
\author[LLNL]{B.~Chung}
\author[UCRiv]{D.E.~MacLaughlin}
\author[UCRiv]{L.~Shu}
\author[UCRiv]{J.E.~Anderson}
\address[LANL]{Los Alamos National Laboratory, MS K764, Los Alamos, NM 87545,
             USA}
\address[LLNL]{Lawrence Livermore National Lab, P.O. Box 808, Livermore, CA 94550 USA}
\address[UCRiv]{Department of Physics, University of California,
              Riverside, CA 92521, USA}
\address[JAERI]{Japan Atomic Energy Research Institute, Tokai-Mura, Naka-Gun, Ibaraki-Ken, 319-1195 Japan}
\address[TRI]{TRIUMF, 4004 Wesbrook Mall, Vancouver, B.C., Canada V6T 2A3}

\begin{abstract}
We present the first $\mu$SR measurements ever performed on elemental Pu, and  set the most stringent upper limits to date on the magnitude of the ordered moment $\mu_{\rm ord}$ in $\alpha$-Pu and $\delta$-stabilized Pu (alloyed with 4.3 at. \% Ga). Assuming a nominal hyperfine coupling field of $1$ kOe/$\mu_{\rm B}$ we find  $\mu_{\rm ord} \leq  10^{-3} \mu_{\rm B}$ at T $\cong 4$~K. 
\end{abstract}

\end{frontmatter}

\section{Introduction}
Understanding the ground-state properties of Pu metal represents a particular challenge for f-electron physics \cite{pu}. This is because the five Pu 5-f electrons sit at the boundary between localization and itinerancy. This ambiguity results in 6 different solid state phases for metallic Pu  as a function of temperature and volume. At room temperature Pu metal exists in the lowest-volume $\alpha$-phase, which possesses a monoclinic structure. At elevated temperatures (near 700 K) Pu reverts to a simpler fcc structure with a greatly expanded unit cell ($\sim +25$~\%), the  $\delta$-phase. This latter phase can also be stabilized at 300 K with the addition of a few percent Ga or Al.  

The large volume increase in the  $\delta$-phase compared to the $\alpha$-phase has been attributed to the partial localization of the f-electrons \cite{wills}. Despite strong evidence for such localization (see, for example, recent photoemisson studies \cite{joyce}), there exits little or no experimental evidence for magnetism in $\delta$-Pu \cite{lashley}. Possible reasons for this could be a crystal-field configuration yielding a non-magnetic ground-state multiplet, Kondo compensation of the moments \cite{Fournier}, as occurs in heavy-fermion systems, approximate cancellation of the orbital and spin moments \cite{orbspin},  as well as other more exotic theoretical possibilities such as noncollinear intra-atomic magnetism \cite{nordstrom}.  A fundamental question concerning these phases, therefore, is to what extent magnetic order, either  ordered or {\it disordered} freezing of the spins, can be completely eliminated. In Kondo lattice systems, for example, small magnetic moments can survive at temperatures much less than the effective Kondo temperature. A recent compilation \cite{lashley} of past neutron (and other) data finds limits of 0.4 - 0.04  $\mu_B$ for the ordered moment in $\delta$-Pu.  These values, while small, are far larger than limits achievable ($\leq 0.001 \mu_B$) with $\mu$SR. Furthermore, $\mu$SR is equally sensitive to the freezing of {\it disordered} spins, and, because $\mu$SR is a local probe, the measurement consists of a sum over all points in momentum space.  Thus,  more stringent upper limits can be set than with most other probes.
 
\section{Experimental results}
$\mu$SR measurements carried out at the M20 surface muon channel at TRIUMF in Vancouver, Canada, for both $\alpha$-Pu and Ga-stabilized $\delta$-Pu (4.3 at. \%~ Ga) in zero field (ZF) and 2.5 kOe  transverse (TF)  field.   The $\alpha$-Pu specimen (99.98\% pure)  was  melted and electro-refined (two years earlier), then electro-mechanically shaped to a disk approximately 12 mm diameter and 0.1mm thick  and  cleaned of oxide. The majority of
the sample was $^{239}$Pu (93.7 \%), with smaller concentrations of $^{240}$Pu (5.86 \%) and $^{238}$Pu (0.17 \%). The specimen contained the following magnetic impurities:  Fe(235 at. ppm), Ni(24 at. ppm), Cr(12.4 at. ppm) and Mn(10.4 at. ppm). The $\delta$-Pu  sample was synthesized from the same Pu as the $\alpha$-Pu, alloyed with Ga and prepared in the same manner.  The magnetic impurity levels were also, within error, the same as the $\alpha$-Pu. 

\begin{figure}
\centering
\includegraphics[width=\columnwidth]{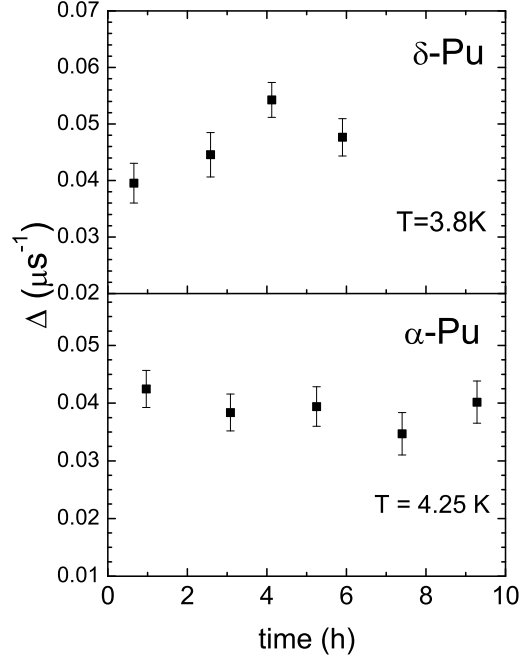}
\caption{
ZF static Gaussian Kubo-Toyabe widths in $\alpha$-Pu and $\delta$-Pu show no significant time dependence as radiation damage accumulates at low temperatures. The times shown are the midpoint times for a particular run.}
\label{timedep}
\end{figure}
\begin{figure}
\centerline{\includegraphics[width=\columnwidth]{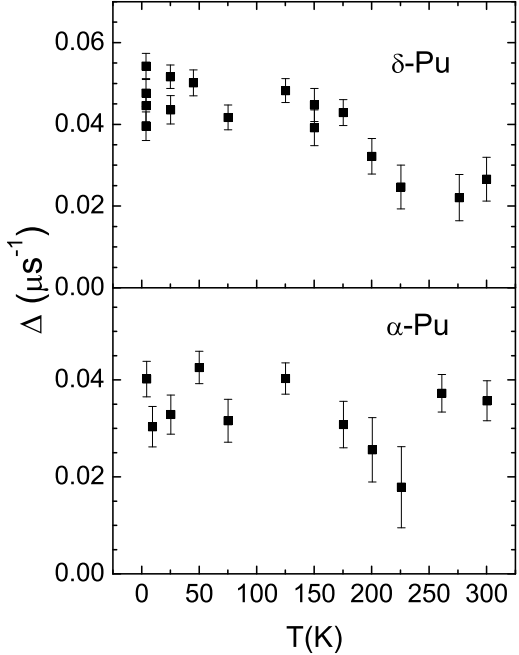}}
\caption{
ZF static Gaussian Kubo-Toyabe widths in $\alpha$- and $\delta$-Pu  are small between $T = 3.8 -300$~K. Muon diffusion may cause the reduction of the linewidth above $200$~K in $\delta$-Pu.}
\label{tempdep}
\end{figure}
The samples were encapsulated in a 70$\mu$m thick Kapton coating and sealed under He atmosphere inside a titanium cell having a 50 $\mu$m Ti-foil beam window and attached to a continuous-flow He cold-finger cryostat. The dual encapsulation was undertaken to prevent possible radioactive contamination.   A negligible fraction of the muons stopped in the Ti
window or the Kapton coating. The background signal obtained from an empty Ti
holder in ZF was well characterized by a static Gaussian Kubo-Toyabe (GKT) relaxation function \cite{Hayano} with rate $\approx 0.014 \mu$s$^{-1}$ at 10K. For ZF experiments the residual field was reduced to $\leq 10$ mOe using trim coils.

The cryostat was mounted along the incident beam direction and the  data  taken with the muon spin rotated approximately 90 degrees vertically from the incoming muon momentum.   The sum of two relaxation functions (for Pu and Ti) was used for fitting : GKT in ZF and Gaussian ($\exp(-\sigma^2t^2/2)$) in TF field.    The rate for Ti was fixed at its measured value, and the relative amplitudes were fixed at a ratio ($\approx$ 2:1, Ti:Pu) determined in a separate experiment with a blank Cu pellet the same size as the Pu samples. The Pu rates for  ZF and TF  are designated $\Delta$ and $\sigma$, respectively. 

Uranium recoil nuclei and alpha particles from the decay of $^{239}$Pu (half-life  $2.4 \times 10^4$ y) damage the lattice by producing interstitials and vacancies. For the $\alpha$- and $\delta$-Pu specimens the displacemnt damage cascade is immobile below about 30K.  For $\alpha$-Pu $>$ 95\% of the low-temperature accumulated damage is annealed at room temperature \cite{Mccall}, while for  $\delta$-Pu the annealing is 100\% at 300 K.  Of course,  U, Am and helium from the Pu decay remain and continue to accumulate.   To check that the muon sensed only undamaged regions, we allowed the samples to sit for up to 9 hours at $T \sim 4$~K, during which time $\mu$SR runs were performed about every 2 hours. No significant change in the $\mu$SR rate was observed over this period, as seen in Fig. \ref{timedep}, indicating no observed cumulative effects from radiation damage. For subsequent data sets, the sample was always re-annealed at $300$~K every 6-8 hours.

\begin{figure}
\centerline{\includegraphics[width=\columnwidth]{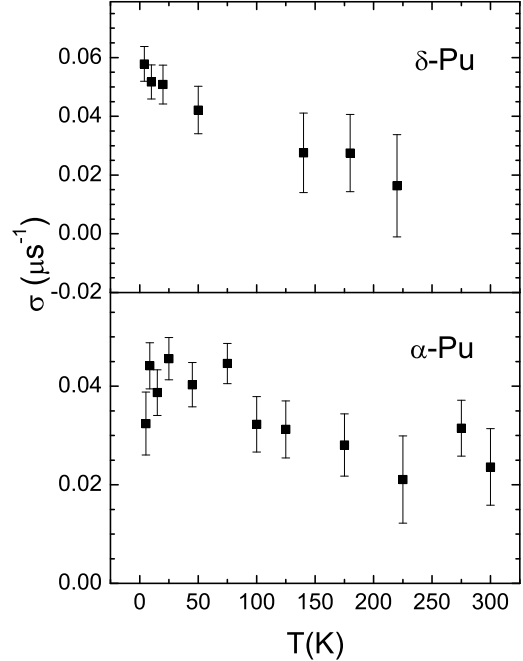}}
\caption{
Temperature dependence of the TF  Gaussian  widths $\sigma$ in $\alpha$- and $\delta$-Pu.}
\label{TF}
\end{figure}

The temperature dependencies of the $\Delta$ and $\sigma$ rates in $\alpha$-Pu and $\delta$-Pu  are shown in Figs. \ref{tempdep} and \ref{TF}. A small longitudinal field ($\sim$ 50 Oe) reduced the ZF $\mu$SR rate to zero, indicating static behavior. In $\alpha$-Pu the data are T-independent. The decreased rates  above $\sim 150$~K in $\delta$-Pu are likely due to muon diffusion causing a site change above this temperature; this will be investigated further in future  experiments.   At all temperatures we find $\sigma \simeq \Delta$, which is expected for a static, random field distribution \cite{Hayano}. A small magnetic field drift ($\ll \sigma$ over a 1 hour run) prevented an accurate analysis of the temperature dependence of the muon frequency.

The muon sites in $\alpha$-Pu are not known. In a perfect fcc $\delta$-Pu  lattice the $\mu^+$ would be expected to reside at the octahedral and/or tetrahedral positions. We calculate  ZF nuclear dipolar linewidths of 0.018 and 0.023 $\mu$s$^{-1}$ for  these sites, respectively. The observed values are  higher, indicating that the muon is perturbed from these high symmetry sites due to the Ga doping or to strain fields.  

We conclude that the small, static, random field distributon and lack of temperature dependence  below $\sim 100$~K  is consistent with nuclear dipolar broadening down to $T \cong 4$~K. The uniform ordering of Pu moments of any significant size ($\sim 0.01 \mu_{\rm B}$) (e.g., ferro- or antiferromagnetism) would be manifest as a precessing $\mu$SR signal below the ordering temperature. Disordered spin freezing would produce a T-dependent, exponential rate \cite{AuFemusr}. Neither is observed.

The absence of a low-temperature $\mu$SR signal from the magnetic impurities is not surprising.  A $\mu$SR rate $\lambda$ of $\approx 5 \mu$s$^{-1}$ is found in the spin glass state of Au{\underline Fe}(1\%)  at $T = 4$~K ($T_{\rm g} = 9.1$~K)  \cite{AuFemusr},  which scales to $\lambda \approx 0.12 \mu$s$^{-1}$ for 235 at. ppm Fe, but only below the estimated \cite{AuFeglass} glass temperature  $T_{\rm g} = 235$ mK. In a similar vein, McCall {\it et al.} \cite{Mccall} have reported an excess, low-temperature, Curie-like susceptibility in $\alpha$-Pu which increases linearly in time, and is interpretated in terms of radiation-induced local moments $\sim 550 \mu_{\rm B}$ per alpha decay.   A crude model indicates  that these moments would not be observed in this experiment either. Assuming a moment $\approx 550 \mu_{\rm B}$  concentrated in the regions of damage, and using Au{\underline Fe} as a template for the spin-glass freezing of such moments ($\mu_{Fe} \approx 2.5 \mu_{\rm B}$), we estimate a freezing temperature of only $7$ mK after a damage accumulation period of 10 hours at $4$ K,  with an average spacing between damage sites of about 530 lattice spacings. Thus, these damage regions should not be (and are not) seen by $\mu$SR.

Our data thus allow us to set a limit on the ordered moment $\mu_{\rm ord}$ in both materials. Using a typical muon hyperfine field  found in f-electron systems of $H_{\rm hyp} \sim 1$~kOe/$\mu_{\rm B}$ and taking $\Delta \simeq 0.05 \mu$s$^{-1}$ as an upper limit for the relaxation rate from the hypothesized ordering, one has   $\mu_{\rm ord} \leq  \gamma_{\mu}/(H_{\rm hyp}~\Delta) \simeq 10^{-3} \mu_{\rm B}$ at $T \cong 4$~K. 

If there is a sizeable localized component to the f-electron density in $\delta$-Pu, why might it  be unobservable? It is possible that the spins do not order until $T \ll 4$~K, so that the mean fluctuation rate $\tau^{-1}$ is still  too rapid  to produce a significant relaxation rate $1/T_1$ in the time scale of our measurement. We can estimate a lower limit for $\tau^{-1}$ by taking $1/T_1 = 2(\gamma_\mu H_{\rm hyp})^2\tau \ll \Delta(4K)$. Then $\tau^{-1}>> 3.6\times10^{11} s^{-1}$, corresponding to an upper limit for the f-electron linewidth $\Gamma = 0.24$ meV, again for $H_{\rm hyp} \sim 1$~kOe/$\mu_{\rm B}$. This is  a few times lower than the $\Gamma$ estimated from NMR measurements in Pu(Ga 1.5 \%) \cite{Piskunov}.

In conclusion, we see no evidence for ordered or disordered magnetic spin freezing in either $\alpha$- or $\delta$-Pu, setting a limit on the frozen moment of less than $0.001 \mu_{\rm B}$.  Whether the moments in $\delta$-Pu are essentially zero, as some recent band structure calculations predict \cite{shick}, are reduced by hybridization \cite{Fournier}, or washed out by dynamical fluctuations \cite{kotliar} is currently unknown. 



We thank the staff of TRIUMF for their kind assistance and E. D. Bauer, L. A. Morales and J. L. Sarrao at Los Alamos for useful discussions.
Work at LANL and LLNL(contract W-7405-Eng-48)  performed  under the auspices of
the U.S. D.O.E.  Work at Riverside supported by the U.S. NSF, Grant DMR-0102293.

\end{document}